\documentclass[lettersize,journal]{IEEEtran}
\usepackage{amsmath,amsfonts}
\usepackage{algorithmic}
\usepackage{algorithm}
\usepackage{array}
\usepackage[caption=false,font=normalsize,labelfont=sf,textfont=sf]{subfig}
\usepackage{textcomp}
\usepackage{stfloats}
\usepackage{url}
\usepackage{verbatim}
\usepackage{graphicx}
\usepackage{multirow} 
\usepackage{cite}
\usepackage{color}

\begin{document}
\title{Efficient Automatic Modulation Classification for Next-Generation Wireless Networks}

\author{To Truong An, ~\IEEEmembership{Student Member,~IEEE}, Antonios Argyriou, Annisa Anggun Puspitasari,\\ Simon L. Cotton, ~\IEEEmembership{Fellow,~IEEE}, and  Byung Moo Lee,~\IEEEmembership{Senior Member,~IEEE}

\thanks{To Truong An, Simon L Cotton are with the Centre for Wireless Innovation (CWI), ECIT Institute, Queen’s University Belfast, Belfast, BT3 9DT, U.K. (e-mail: ato01@qub.ac.uk, simon.cotton@qub.ac.uk)

Annisa Anggun Puspitasari, Byung Moo
Lee are with the Department of Artificial Intelligence and Robotics,
and Convergence Engineering for Intelligent Drone
, Sejong University, Seoul, 05006, South Korea, (e-mail: annisanggun@gmail.com, blee@sejong.ac.kr).

Antonios Argyriou is with the Department of Electrical and Computer
Engineering, University of Thessaly, Volos, Thessaly 38221, Greece (e-mail:
anargyr@uth.gr)

This work was supported in part by the Cyber AI Hub Doctoral Training Programme at CSIT, Queen’s University Belfast, by the UK Government as part of the New Deal for Northern Ireland, administered through Innovate UK/UKRI; in part by the Basic Science Research Program through the National Research Foundation of Korea (NRF) funded
by the Korea Government (MSIT) under Grant NRF-2023R1A2C1002656; in part by the Institute
of Information \& Communications Technology Planning \& Evaluation (IITP)-
ICT Challenge and Advanced Network of HRD (ICAN) Grant funded by the
Korea Government (Ministry of Science and ICT) under Grant IITP-2025-RS-2022-00156345; and in part by the Ministry of Science and ICT (MSIT), South
Korea, under the Information Technology Research Center (ITRC) support
program supervised by the Institute for Information \& Communications
Technology Planning \& Evaluation (IITP) under Grant IITP-2025-RS-2024-
00437494.
}}
   
\markboth{IEEE Transactions on Green Communications and Networking}%
{Shell \MakeLowercase{\textit{et al.}}: A Sample Article Using IEEEtran.cls for IEEE Journals}


\maketitle

\begin{abstract}
%This paper introduces a novel automatic modulation classification (AMC) algorithm for wireless communication systems. 
With the imminent development of sixth-generation (6G) networks, there will be a demand for high-accuracy, computationally-efficient, and low-inference time automatic modulation classification (AMC) algorithms. To address this need, we propose a new deep-learning based model for AMC that is called the threshold denoise recurrent neural network (TDRNN). The TDRNN combines an adaptive threshold denoising (TD) algorithm and a recurrent neural network (RNN) that together achieve high accuracy and fast inference. The TD module adaptively reduces the noise level of the received signal, while the RNN module performs the modulation classification on the denoised result. The two subsystems are jointly optimized to reach the optimal architecture. The proposed TDRNN is evaluated for various modulation schemes and signal-to-noise ratios (SNR). The experimental results demonstrate that the TDRNN outperforms existing methods in terms of accuracy, speed, and computational complexity making it an ideal solution for 6G wireless communication systems. 
\end{abstract}


\section{INTRODUCTION}
\IEEEPARstart{I}{n} the rapidly evolving landscape of mobile wireless networks, successive generation cellular systems are introduced approximately once every decade driven in part by the demand for enhanced data rates and improved wireless connectivity. With the successful launch of fifth-generation (5G) wireless communication systems, the groundwork has been laid for the exploration and development of sixth-generation (6G) systems \cite{ps7}. Although 6G is still in its early stages, several key performance indicators (KPIs) have already been identified as the focal points that will drive specific technical decisions \cite{b9,b21}.

One of the envisioned KPIs for 6G systems includes ultra-high reliability where the aim is to ensure that the wireless link is available 99.99999\% of time \cite{b18}. This highly ambitious target is needed to enable new highly dependable and robust communication services across a variety of mission-critical scenarios. Ultra-low latency is another critical requirement for 6G networks, where network delays less than 0.1 milliseconds (ms) will be targeted. Extremely low latency like this ensures near-instantaneous transmission and response times, setting the stage for seamless and real-time applications \cite{b9}.

Against this backdrop, future wireless systems will require supporting algorithms that will need to be heavily optimized to ensure their low-latency operation. One class of crucial algorithms for future wireless communication is automatic modulation classification (AMC). Principally, AMC will primarily be designed to identify the modulation of unknown wireless signals. In the context of 6G systems, AMC will be a key technology since modulation classification is critical for optimizing spectral efficiency \cite{b1}, mitigating interference \cite{b3}, enabling channel-adaptive communication, enhancing security, facilitating cognitive radio and spectrum sharing, streamlining signal processing at the receiver \cite{b2}, and elevating the quality of service \cite{b1}. As communication systems become more complex, AMC’s ability to classify modulation schemes efficiently will play a crucial role in enhancing overall system performance in future wireless networks \cite{b2,b3}.

\subsection{AMC in The Deep Learning Era}
While traditional AMC approaches can be categorized as likelihood-based (LB) and feature-based (FB), in recent years, machine learning (ML)-based schemes have dominated. ML, and particularly deep learning (DL), has emerged as a prominent technology across various engineering and scientific domains. The efficacy of DL in real-time decision-making has led to its widespread application in adaptive and challenging wireless environments \cite{b8}. DL has also been successfully applied in AMC, where artificial neural networks (ANN), convolutional neural networks (CNN)\cite{ps9}, and recurrent neural networks (RNN) have been widely used \cite{b14, ps8}.

CNNs, specifically designed for image and video processing tasks, excel at automatically learning spatial feature hierarchies from input images or video frames. This makes them highly suitable for AMC scenarios involving visual information (e.g. spectrograms). On the other hand, RNNs are well-suited for processing sequential data, such as natural language or time series data. Thus, RNNs have shown their superiority when processing the in-phase and quadrature components of complex baseband communication signal samples \cite{ps10}.

In the context of developing an AMC system for ultra-reliable low-latency communications (URLLC) in 6G networks, RNNs have emerged as a promising technology. RNNs possess inherent capabilities to process sequential data and capture temporal dependencies, making them well-suited for modulation classification. Due to their recurrent connections, RNNs can efficiently learn and adapt to evolving modulation schemes, enabling faster and more accurate modulation classification. This speed and adaptability position RNNs as one of the first choices for meeting the stringent requirements of 6G networks. In terms of latency, the performance of the AMC RNN-based model is expected to be highly efficient. 
\subsection{This Paper}
This paper presents a novel approach to AMC in wireless communication, underpinned by the \textit{Threshold Denoise Recurrent Neural Network (TDRNN)} model. The primary objective of our approach is to vastly improve model size and inference time while at the same time delivering classification performance that is comparable to the state-of-the-art. The proposed model combines two crucial components: The threshold denoiser (TD) layer and the recurrent neural network (RNN) layer. The first component is responsible for denoising the input signal based on a threshold which is calculated dynamically depending on the modulation and the noise conditions. After denoising the input signals, they are classified by the RNN. 

The system that we propose, and briefly described, aims to enhance both the speed and accuracy of AMC in emerging wireless communication systems that require low latency in AMC inference time.

Overall, the contributions of this work are the following: 
\begin{itemize}
  \item We propose a novel approach to AMC, based upon TDRNN, specifically targeted at delivering URLLC in future networks such as 6G. Our model achieves an impressive inference time of 0.007 ms per a single modulation example, while reducing computational overhead and energy consumption. The proposed scheme outperforms state-of-the-art methods in both inference time and classification accuracy, making it well-suited for green and sustainable wireless communication networks.
  \item We demonstrate the effectiveness of the Threshold Denoiser, an adaptive algorithm that automatically learns the optimal denoising threshold for each modulation scheme based on the SNR. This adaptive approach significantly reduces the impact of noise, leading to a remarkable improvement of nearly 16\% in the average accuracy of the AMC model.
  \item We conduct an in-depth study of the architecture of the TDRNN, and analyze in detail the impact of the number of Gated Recurrent Units (GRU) in the TDRNN model, as well as the influence of the number of hidden layers within the GRU. This means that our results can be readily used to design the optimal TDRNN model, tensioning performance against complexity.
  \item We assess the performance of the TDRNN through a comprehensive evaluation, comparing it with state-of-the-art AMC techniques. Our model demonstrates significant improvements in three key areas: it has the fewest parameters, approximately 2 to 10 times fewer than the benchmark; it achieves the fastest inference speed, with a time of only 0.007 ms per a single modulation example; and it provides competitive classification accuracy, outperforming state-of-the-art methods on the RadioML 2016.10A dataset and delivering comparable performance on the RadioML 2018.01A dataset.

\end{itemize}

\subsection{Paper Organization}
The rest of this paper is organized as follows. Section II provides an overview of the related work, while Section III presents in detail our assumptions and the proposed system. Section IV presents the results and discusses the thorough performance evaluation and comparison with related work. Finally, Section V concludes the paper by summarizing the key findings and their implications of wireless communication systems. 

\section{RELATED WORK}
\label{sec: related work} 
AMC based on DL has been investigated in several studies in recent years due to its promise to enhance the performance of future networks.  In \cite{rw1} an AMC system based on a CNN model that consisted of two convolutional layers and three fully-connected layers was proposed. The system was designed for orthogonal frequency division multiplexing (OFDM), and also considered the presence of phase offset in the signal. Even though the system delivered a significant accuracy improvement in the classification of several modulation techniques, it was found to be very complex. While in \cite{rw2}, the authors proposed a Complex CNN model to compute a complex convolution. The problem was solved by implementing a linear combination that has a two-dimensional in-phase and quadrature (I/Q) data stream. While this system provided a 30\% improvement for a single I/Q model with 256 and 260 nodes, but required 2.7 million parameters, which is higher than the number of parameters required by the approach proposed here. 

Another study succeeded in reducing the number of parameters required by implementing a system called convolutional, long short-term memory, deep neural network (CLDNN) \cite{rw3}. This system combined shallow CNNs and long short-term memory (LSTM) and used a course-to-fine method to find the best length for the input sequences. The results showed that the accuracy in the time domain, frequency domain, or autocorrelation domain was increased. Based on the comparison in \cite{rw4}, the CLDNN required 849K parameters. On the other hand, a system named MCLDNN implemented integrated one-dimensional convolution, two-dimensional convolution, and RNN classification \cite{rw5}. MCLDNN adopted a multi-stream structure to process information from multiple I/Q channels, equipped with independent input channels to extract signal features. As a result of this approach, both in-phase and quadrature spatial features can be combined, where the CNN was responsible for spatial mapping, and the LSTM was applied to analyze the sequential data by considering it as part of the temporal extraction procedure. This resulted in up to 10\% accuracy improvement over previous methods, while MCLDNN required 406K parameters. Overall, these studies have shown that while DL can be complex, in terms of the number of parameters required for model implementation, they show much promise for modulation type identification. In this instance, DL is learning to identify complex patterns that are difficult to detect with traditional model-based methods.

When there is a high number of parameters, this can lead to a rise in computational complexity. This, in turn, can have a negative impact on the overall effectiveness of the system. The authors in \cite{rw6} presented the SCRNN that reduces the number of parameters required and does not require hand-crafted expert features. The system used a CNN for feature extraction and dimensionality reduction and also an LSTM to remember long-term dependencies. Requiring 398K parameters, the system improved identification accuracy by up to 15\% and reduced training time by 74\% compared to independent CNN or RNN systems. 

\begin{table*}[t]
\centering
\caption{Related previous works.}
\setlength{\tabcolsep}{12pt}
\renewcommand{\arraystretch}{1.5}
\label{tab-rw}
%\resizebox{\columnwidth}{!}{%
\begin{tabular}{||c|c|c|c|c||}
\hline \hline
\multicolumn{1}{||c|}{\textbf{References}} &
  \multicolumn{1}{c|}{\textbf{Denoise}} &
  \multicolumn{1}{c|}{\textbf{Basic Structure}} &
  \multicolumn{1}{c|}{\textbf{Parameters}} &
   \multicolumn{1}{c||}{\textbf{Inference time (ms)}} \\ \hline \hline
CNN-Based AMR {\cite{rw1}}                                                                     & x          & CNN                & 4 million       & not mentioned       \\ \hline
Complex CNN {\cite{rw2}}                                                                       & x          & CNN                & 2.7 million     & not mentioned       \\ \hline
CLDNN {\cite{rw3}}                                                                              & x          & CNN + LSTM         & 849,583         & 0.477               \\ \hline
MCLDNN {\cite{rw5}}                                                                             & x          & RNN                & 406,199         & 0.061               \\ \hline
SCRNN {\cite{rw6}}                                                                              & x          & CNNs + LSTMs       & 398,731         & 0.661               \\ \hline
CGDNet {\cite{rw4}}                                                                            & x          & CNN + RNN          & 124,676         & 0,267               \\ \hline
TL-CSNN {\cite{rw8}}                                                                            & x          & DNN + TL           & 119,627         & not mentioned       \\ \hline
SCNN {\cite{rw9}}                                                                              & x          & CNN                & 96,020          & not mentioned       \\ \hline
Combined CNN {\cite{rw10}}                                                                      & x          & Deep + Shallow CNN & 113,000         & not mentioned       \\ \hline
PET-CGDNN {\cite{rw13}}                                                                         & x          & CNN                & 71,871          & 0.039               \\ \hline
MCMBNN {\cite{rw14}}                                                                            & x          & CNN                & 63,494          & not mentioned       \\ \hline
\textbf{TDRNN (Ours)}                                                               & \textbf{\checkmark} & \textbf{RNN}       & \textbf{41,821} & \textbf{0.007}      \\ \hline \hline
\end{tabular}
\end{table*}

Another interesting study proposed a deep learning-based model called CGDNet, which consists of three convolutional layers followed by an LSTM block to detect temporal changes in the given modulation sequence \cite{rw4}. The system underwent comparative analysis with various state-of-art models, such as the CLDNN model. The result found that the CGDNet model exhibited an improvement in accuracy of up to 36.6\% compared to prior research. When compared with the CLDNN model, it was found that CGDNet required significantly fewer parameters, standing at 124K. In addition, CGDNet boasted a faster inference time of 0.21 ms, 85\% faster than the CLDNN model. TL-CSNN proposed in \cite{rw8} consists of a cascaded single neural network with transfer learning that identifies the modulation format and optical signal-to-noise ratio (SNR). Because the authors implemented transfer learning, they used identical training sets for all schemes to ensure a fair comparison while comparing their system with other models. The experimental results indicated that with 119K parameters, the proposed approach demonstrated faster convergence, greater precision, heightened stability, and with down to 37\% fewer parameters that previous models. In \cite{rw9}, the authors focused on implementing decentralized learning for AMC with a separable CNN (SCNN). SCNN featured model consolidation and lightweight design. The proposed model proved to be more efficient than the SCNN-based centralized AMC. It enhanced training efficiency and reduced communication overhead while maintaining classification performance. With 96K parameters, the training efficiency of SCNN was determined as being approximately N times that of SCNN-based centralized learning, where N represented the number of edge devices. Their model showcased enhanced accuracy compared to standard CNN while reducing space and time complexity by as much as 94\%  and 96\%, respectively. 

According to the aforementioned studies, fewer parameters are required for system training. However, it is important to acknowledge that the decline in the number of parameters is related to the level of accuracy, as shown in \cite{rw10}. Consequently, it is important to consider all relevant factors so as to ensure optimal system performance.

To address this, several studies have tried to enhance the accuracy while reducing the number of parameters and decreasing the inference time. 
In \cite{rw11}, the authors proposed a combined CNN scheme to identify single modulation mode signals accurately. To overcome the challenges of misclassifying single-sideband amplitude modulation with carrier and suppressed carrier, the authors employed a shallow network to complement both of those networks, which have a deep structure. Utilizing 113K parameters, their system achieved a peak accuracy of 98.7\%.

A different research study introduced a system known as PET-CGDNN \cite{rw13}. This system prioritized lightweight and low-complexity models by reducing the kernel size of CNN layers and that of feature maps. Furthermore, it implemented a parameter estimator and transformer to minimize the negative impact on the phase of the signal. The authors conducted a thorough analysis by comparing PET-CGDNN with other models, including MCLDNN. It was found that the PET-CGDNN model exhibited an average accuracy of 60.44\% while utilizing fewer parameters and computational resources. Meanwhile, in \cite{rw14}, a lightweight decentralized-learning-based AMC system called MCMBNN was proposed. This system utilized spatiotemporal hybrid DNN based on multichannel and multifunction blocks. The authors applied three channels to extract the phase, spatial, and temporal features with low complexity in certain function blocks. Results indicated that the system required 63K parameters to train the RML2016.10A dataset with GRU 64 delivering a 55.82\% probability of correct classification. Based on the combined CNN scheme \cite{rw11}.
\begin{figure*}[!t]
\centering
\includegraphics[width=6in]{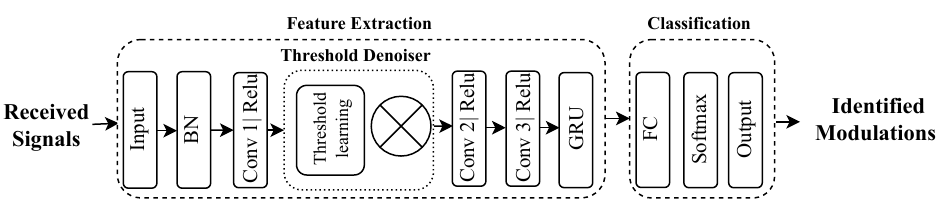}
\caption{Proposal TDRNN architecture.}
\label{fig1}
\end{figure*}
\begin{figure*}[!t]
\centering
\includegraphics[width=6in]{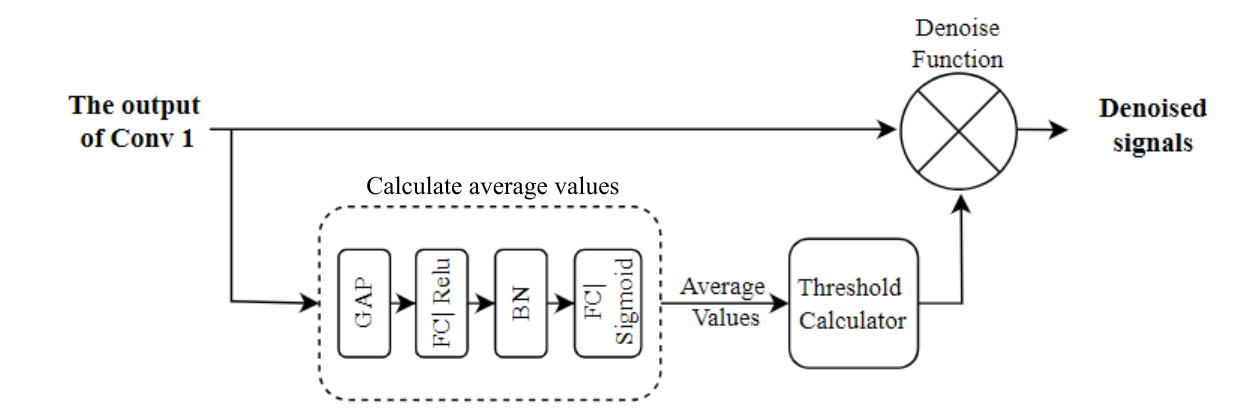}
\caption{Threshold Denoiser architecture.}
\label{fig2}
\end{figure*} 

As highlighted in the aforementioned studies, it is apparent that high accuracy could be accomplished even with a few parameters. However, in the process of designing AMC, it is crucial to consider not only accuracy and parameter count but also the computational cost that translates to latency in inference time. The latter is particularly important in the context of future wireless networks (e.g.6G), which is expected to achieve the URLLC. Our system takes a significant step towards achieving this goal, proposing a powerful new AMC approach that delivers high accuracy, achieved with low computational cost. A comparative summary of our system with the literature models that we thoroughly discussed is presented in Table \ref{tab-rw}.

\section{SIGNAL MODEL AND METHOD}
\subsection{Signal Model}

This work considers modulation classification in real wireless channel conditions, incorporating effects such as frequency-selective fading, delay spread, sampling timing offset (STO), carrier frequency offset (CFO), additive white Gaussian noise (AWGN), etc.

The received baseband modulated signal $r(t)$ is: 
\begin{equation}
    r(t) = A(t)e^{j(2\pi f_{o} t+\theta)} \sum_{k } s[k]h(t-kT+\nu) + n(t),
\end{equation}
where $n(t)$ represents the complex AWGN, and $A(t)$ is the amplitude of the received signal at time $t$. Regarding the phase of the signal in the exponential term, it is affected by $f_{o}$ which is the CFO that creates a time varying phase offset, and $\theta$ indicates the constant phase offset of the signal. In this expression we sum the output of each transmitted symbol $s[k]$ (indexed by $k$) with $T$ being the symbol period. The impulse response $h(.)$ contains the combined effect of transmit and receive filtering (e.g. raised cosine) and the channel. Finally, $\nu$ indicates the STO.

The received signal $r(t)$ is sampled with an analog-to-digital converter (ADC) that brings it to a discrete form. Sampling with rate $f_s$ Hz, the in-phase and quadrature (IQ) components of the complex signal $r(t)$, produces complex samples $r[n]$. %The IQ samples can be used to determine the frequency and phase of the signal~\cite{ps2}. 

The complex samples can be decomposed into its discrete IQ components: $I\left[ n \right] = \Re \left( r\left[  n \right] \right)$ with $\Re$ being the operator for real part. Similarly, $Q\left[ n \right] = \Im \left( r\left[ n \right] \right)$, with $\Im$ being the operator for imaginary part. In this study, we use I/Q information as an input to the AMC model. The input signal vector can be described as follows:
\begin{equation}
    X =\binom{X_{i}}{X_{q}}= \binom{\Re \left[ r[1], .....r[L] \right]}{\Im \left[ r[1], .....r[L] \right]}
\end{equation}
where $X$ is a complex vector composed of two parts which are $X_{i},X_{q}$ denoting the in-phase and quadrature components respectively, $r[.]$ is the received complex baseband sample, and $L$ is the length of the sample that will be used by the TDRNN.

\subsection{Proposed Method}

\begin{table*}[t]
\centering
\caption{TDRNN architecture}
\setlength{\tabcolsep}{12pt}
\renewcommand{\arraystretch}{1.5}
\label{table 3}
%\resizebox{1.6\columnwidth}{!}{%
\begin{tabular}{lccl}
 \hline \hline
\multicolumn{1}{||c|}{\textbf{Layer Type}} &
  \multicolumn{1}{c|}{\textbf{Input Size}} &
  \multicolumn{1}{c|}{\textbf{Output Size}} &
   \multicolumn{1}{c||}{\textbf{Details}} \\ \hline \hline
\multicolumn{1}{||c|}{Input} &
  \multicolumn{1}{c|}{2x128x1} &
  \multicolumn{1}{c|}{-} &
  \multicolumn{1}{c||}{-} \\
\multicolumn{1}{||c|}{BN} &
  \multicolumn{1}{c|}{2x128x1} &
  \multicolumn{1}{c|}{2x128x1} &
  \multicolumn{1}{c||}{-} \\
\multicolumn{1}{||c|}{Conv2D} &
  \multicolumn{1}{c|}{2x128x1} &
  \multicolumn{1}{c|}{2x128x16} &
  \multicolumn{1}{l||}{ Filter: 16, Kernel size : 2 x 3, ReLu} \\
\multicolumn{1}{||c|}{Auto Learning Threshold} &
  \multicolumn{1}{c|}{2x128x16} &
  \multicolumn{1}{c|}{2x128x16} &
  \multicolumn{1}{c||}{-} \\
\multicolumn{1}{||c|}{Conv2D} &
  \multicolumn{1}{c|}{2x128x16} &
  \multicolumn{1}{c|}{2x128x32} &
  \multicolumn{1}{l||}{Filter: 32, Kernel size : 2 x 3,ReLu} \\
\multicolumn{1}{||c|}{Conv2D} &
  \multicolumn{1}{c|}{2x128x32} &
  \multicolumn{1}{c|}{1x126x64} &
  \multicolumn{1}{l||}{Filter: 64, Kernel size : 2 x 3,ReLu} \\
\multicolumn{1}{||c|}{Reshape} &
  \multicolumn{1}{c|}{1x126x64} &
  \multicolumn{1}{c|}{126x64} &
  \multicolumn{1}{c||}{-} \\
\multicolumn{1}{||c|}{GRU} &
  \multicolumn{1}{c|}{126x64} &
  \multicolumn{1}{c|}{64} &
  \multicolumn{1}{l||}{Hidden units: 64} \\
\multicolumn{1}{||c|}{FC} &
  \multicolumn{1}{c|}{64} &
  \multicolumn{1}{c|}{11} &
  \multicolumn{1}{l||}{Softmax} \\ \hline \hline
\multicolumn{4}{l}{*The architecture has 41,821 parameters}                  
\end{tabular}%
%}
\end{table*}
\begin{figure}[t]
\centering
\includegraphics[width=3.4in]{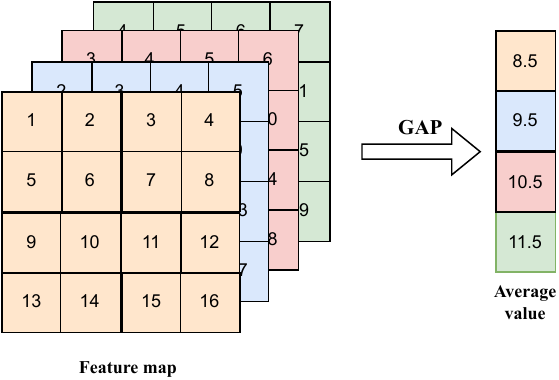}
\caption{An illustration of how global average pooling operates.}
\label{Global Average Pooling}
\end{figure}
In this part, we present the proposed novel architecture for a deep-learning network based on a RNN \cite{8053243} and adaptive denoising. This model is called Threshold Denoise Recurrent Neural Network (TDRNN), which discovers modulation patterns automatically in the training stage and determines their class in the prediction stage. 

The predicted modulation type is the one that maximizes the posterior probability value for the current input signal, expressed as:
\begin{equation}
   y_\text{pred}= \arg \max_{y} f(y|X;W),
    \label{eq1}
\end{equation}
where $y$, $y_\text{pred}$ are the ground truth and the predicted modulation type, respectively, $X$ is the current input signal, and $W$ represents the weights of the model.

The TDRNN model consists of two main parts: Feature Extraction and Classification. Fig. \ref{fig1} shows the architecture of the TDRNN model. 

\subsubsection{Feature Extraction}  

The TDRNN starts with the input layer, in which the sample size is 2 $\times L$. In this study, we set $L = 128$ to match the size of the frames in the RadioML 2016.10A dataset but of course our model can support different $L$. After the input layer, batch normalization (BN) is used to reduce covariance shift and normalize the feature in every sample. BN makes the following calculations:
\begin{equation}
\mu=\frac{1}{k} \sum_{n=1}^{k}x_{n}
\end{equation}
\begin{equation}
\sigma^{2}=\frac{1}{k}\sum_{n=1}^{k}(x_{n}-\mu)^{2}
\end{equation}
\begin{equation}
\widehat{x}_{n}=\frac{x_{n}-\mu}{\sqrt{\sigma^{2}+\varepsilon}}
\end{equation}
where $x_{n}$ and $\widehat{x}_{n}$ represent the input and the normalization input features of each observation in a mini-batch, respectively.  $k$ represents neurons in this layer, $\mu$ is the mean, $\sigma^{2}$ indicates its standard deviation, while $\varepsilon$ is a small constant with $\epsilon\rightarrow 0$.

\begin{table}[t]
\centering
\caption{Influence of the number of GRU layers on model performance}
\renewcommand{\arraystretch}{1.75}
\label{Influence of the Number of GRU Layers on Model Performance}
\resizebox{\columnwidth}{!}{%
\begin{tabular}{||c|c|c|c||}
\hline \hline
\textbf{GRU layer} & \textbf{Parameters} & \textbf{Interence time (ms)} & \textbf{Average accuracy \%} \\ \hline \hline
\textbf{1 layer} & 41821 & 0.0074 & 0.635 \\ \hline
\textbf{2 layer} & 66781 & 0.01   & 0.63  \\ \hline
\textbf{3 layer} & 91741 & 0.012  & 0.634 \\ \hline \hline
\end{tabular}%
}
\end{table}

As soon as the signal has been normalized, it is passed into the first convolutional layer, which contains 16 filters. This layer uses rectified linear units (ReLU) for activation: 
\begin{equation}
ReLU(x) = 
\begin{cases} 
1, &\mbox{if $ x > 0$},\\
0, &\mbox{if $ x \le 0$}.
 \end{cases}
 \label{ReLu}
\end{equation}

In the next step, the signal is denoised by the threshold denoiser sub-network, which can automatically learn the threshold value for each incoming signal. The threshold denoiser structure is illustrated in Fig \ref{fig2}. 

Since noise levels vary across different modulation types, each modulation scheme encounters a distinct level of noise. As a result, a fixed threshold would be insufficient for effective denoising. To address this, we propose a dynamic denoising function that automatically adjusts the threshold based on the noise characteristics present in the input signal. This adaptive approach ensures that only noise-related components are removed while preserving essential modulation features, thereby enhancing classification performance.

For calculating the average value of the signal, we apply the global average pooling (GAP) operation in order to get an average value of the feature map. The GAP can be described graphically with the help of Fig. \ref{Global Average Pooling}. Once we have the average value of the feature map, this value is passed to two fully connected (FC) layers and one Sigmoid function to obtain the average value of the signal. The Sigmoid function can be expressed as follows:

\begin{equation}
   \text{Sigmoid}(x)=\frac{1}{1+e^{-x}}
    \label{Sigmoid}
\end{equation}
After obtaining the average values, the threshold $\tau$ is calculated based on the average values of a signal as: 
\begin{equation}
    \tau = 2\beta
    \label{eq4}
\end{equation}
where $\beta$ is the average value of the signal. 

The signal is denoised by transforming features that are close to zero into zeros, effectively eliminating noise-related features. Formally, the denoising operation can be described as follows:
\begin{equation}
y_{i} = 
\begin{cases} 
0, &\mbox{if $ |x| \le  \tau$}\\
Sgn(x)(|x|-\tau), &\mbox{if $ |x| > \tau$}.
 \end{cases}
\end{equation}
The input and output features are represented by $x$ and $y_{i}$, respectively, and $\tau$ is the threshold value. $Sgn(x)$ is the sign function, which can be expressed as follows:

\begin{equation}
Sgn (x): = 
\begin{cases} 
1, &\mbox{if $ x > 0$}\\
0, &\mbox{if $ x = 0$}\\
-1, &\mbox{if $ x < 0$}.
 \end{cases}
\end{equation}

After denoising, the signal is ready for classification. We use two convolutional layers with ReLU activation and a GRU layer with 64 hidden units to discover the modulation pattern. It is important to note that the convolution layer produces a 4D tensor as an output. Thus, we use a \textit{Reshape layer} to change the input shape to the expectation of the GRU, where the 3D tensor is the input. Table 2 shows details about our model. 

\subsubsection{Classification}
The TDRNN uses a fully connected layer with 11 units, representing the 11 modulation types of the dataset. %, for modulation type classification. 
A Softmax activation is employed to obtain probability distributions over the classes. The Softmax activation normalizes the output from previous layers, ensuring that the predicted probabilities sum up to 1. This facilitates confident predictions and enables the identification of the most probable class for a given input. The Softmax activation can be mathematically expressed as follows for the $i$-th class:
\begin{equation}
    \text{Softmax}(x)_i = \frac{e^{x_{i}}}{\sum_{j}^{\text{Classes}}e^{x_{j}}}
    \label{eq5}
\end{equation}

In addition, for the optimization process, we use the Adam optimizer along with the categorical cross-entropy (CCE) function as the loss function. The CCE function is defined as:
\begin{equation}
L_\text{CCE}=\sum_{i=1}^{M} Y_{1}\log(Y_{2})
\end{equation}
\begin{equation}
Y_{2}=f_{\text{Softmax}(x)_{i}}
\end{equation}
In the above $Y_{1}$ represents the ground truth vector, which can be encoded using one-hot encoding. $Y_{2}$ indicates the predicted vector. $M$ refers to sample types, and $\text{Softmax}(x)_{i}$ corresponds to the result of the $i^{th}$ AMC output.

\subsection{Dataset}
In this work, we employ two widely used datasets, RadioML 2016.10A, and RadioML 2018.01A, which simulate realistic wireless communication scenarios with various modulation schemes and channel effects, including additive white Gaussian noise, multipath fading, frequency offsets, and timing offsets \cite{o2016convolutional, roy2018over}. 
\begin{itemize}
  \item The RadioML 2016.10A dataset contains 220K signals for 20 different SNRs from -20dB to +18dB with a frame length of 128 samples. This equates to 1K signals per modulation technique per SNR level. It consists of 11 modulation schemes: AM-DSB, AM-SSB, BPSK, QPSK, 8PSK, QAM16, QAM64, GFSK, CPFSK, PAM4, and WBFM \cite{o2016convolutional}.
  \item The RML2018.01A dataset contains more than 2.5 million signals for 26 different SNRs from -20dB to +30dB with a frame length of 1024 samples. This dataset includes 24 modulation schemes: OOK, ASK4, ASK8, BPSK, QPSK, PSK8, PSK16, PSK32, APSK16, APSK32, APSK64, APSK128, QAM16, QAM32, QAM64, QAM128, QAM256, AMSSBWC, AMSSBSC, AMDSBWC, AMDSBSC, FM, GMSK and OQPS \cite{roy2018over}.
\end{itemize}
Due to hardware limitations, we select half of the RML2018.01A dataset, ensuring that all modulation schemes are represented with an equal number of samples.
\subsection{Training and Testing}  
%In this section, we introduce a methodology for evaluating the performance of TDRNN using the RadioML 2016.10A dataset. 
This paper uses the RadioML 2016.10A for training and evaluating our proposed model. This is because the RadioML 2016.10A is the most challenging dataset. For this dataset, the state-of-the-art model, which is MCLDNN \cite{rw5} achieves the highest accuracy of only 91\%. However, the highest accuracy achieved by the MCLDNN for the more comprehensive RadioML 2018.01A dataset is 97\%. 

While RadioML 2018.01A contains way more modulations, it is considerably larger and increases significantly the training time. Hence, the benefit of RadioML 2016.10A is that it is relatively smaller by more than 30 times, thereby reducing training costs.

To train the TDRNN model, we employ a random selection approach, allocating 60\% of the data for training, 20\% for testing, and 20\% for validation. For training, we set the hyper-parameters to 500 epochs, the learning rate to 0.001, and used 1024 batches. To enhance the learning process and prevent overfitting, we incorporate an Early Stopping procedure and use the Adam optimization algorithm. The TDCNN model is built using Google's TensorFlow 2.9, a machine-learning framework based in Python. Our experiments and models utilize Keras, TensorFlow, and NVIDIA GeForce RTX 2080 GPUs.

\subsection{Evaluation Metrics}
To evaluate the performance of the proposed system, we consider two key metrics: inference time and average accuracy. These metrics provide insights into both the efficiency and effectiveness of our approach.

Inference time is a key metric for assessing latency in real-time signal classification. It represents the average time required for the deep learning model to classify a single modulation example. Inference time is defined as:
\begin{equation}
    T_{inf} = \frac{T_{total}}{N}
\end{equation}
where $T_{inf}$ is the inference time to classify a single modulation example,  $T_{total}$ is the total inference time for all modulation examples in the test dataset, and $N$ is the total number of modulation examples in the test set.

Average accuracy represents the overall classification performance, evaluated across different modulation schemes and SNR levels. It can be expressed as:

\begin{equation}
\overline{Acc} = \frac{\sum_{_{}}^{}Accuracy}{N_m}
\end{equation}
where $\overline{Acc}$ is average accuracy across all modulations and SNR values, $\sum_{_{}}^{}Accuracy$ is sum of accuracy values for all modulations at all SNR levels, $N_m$ is the total number of modulation schemes evaluated across all SNR levels.

\section{EXPERIMENTAL RESULTS AND DISCUSSION}
In this section, we first conduct some experiments in order to determine the impact each component has on the overall system in the proposed model, and then, based on the results of the experiments, we choose the ideal architectural design. Second, we conduct a comparative analysis of our model against current work across two datasets. Finally, we discuss our findings. 

\subsection{ Experimental Evaluation}  
\begin{figure}[!t]
\centering
\includegraphics[width=3.3in]{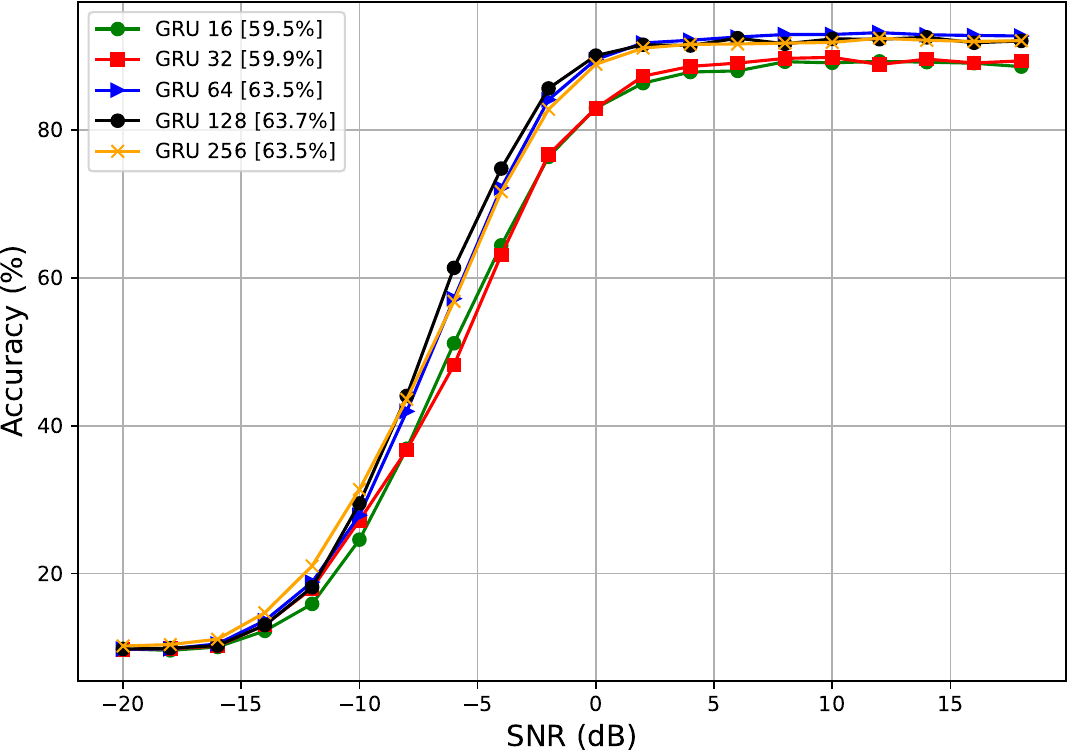}
\caption{Classification performance of different GRU hidden layers. The average classification accuracy of each model is presented within ``[.]". }
\label{Classification Performance of different GRU}
\end{figure}

\begin{figure}[!t]
\centering
\includegraphics[ width=3.6 in]{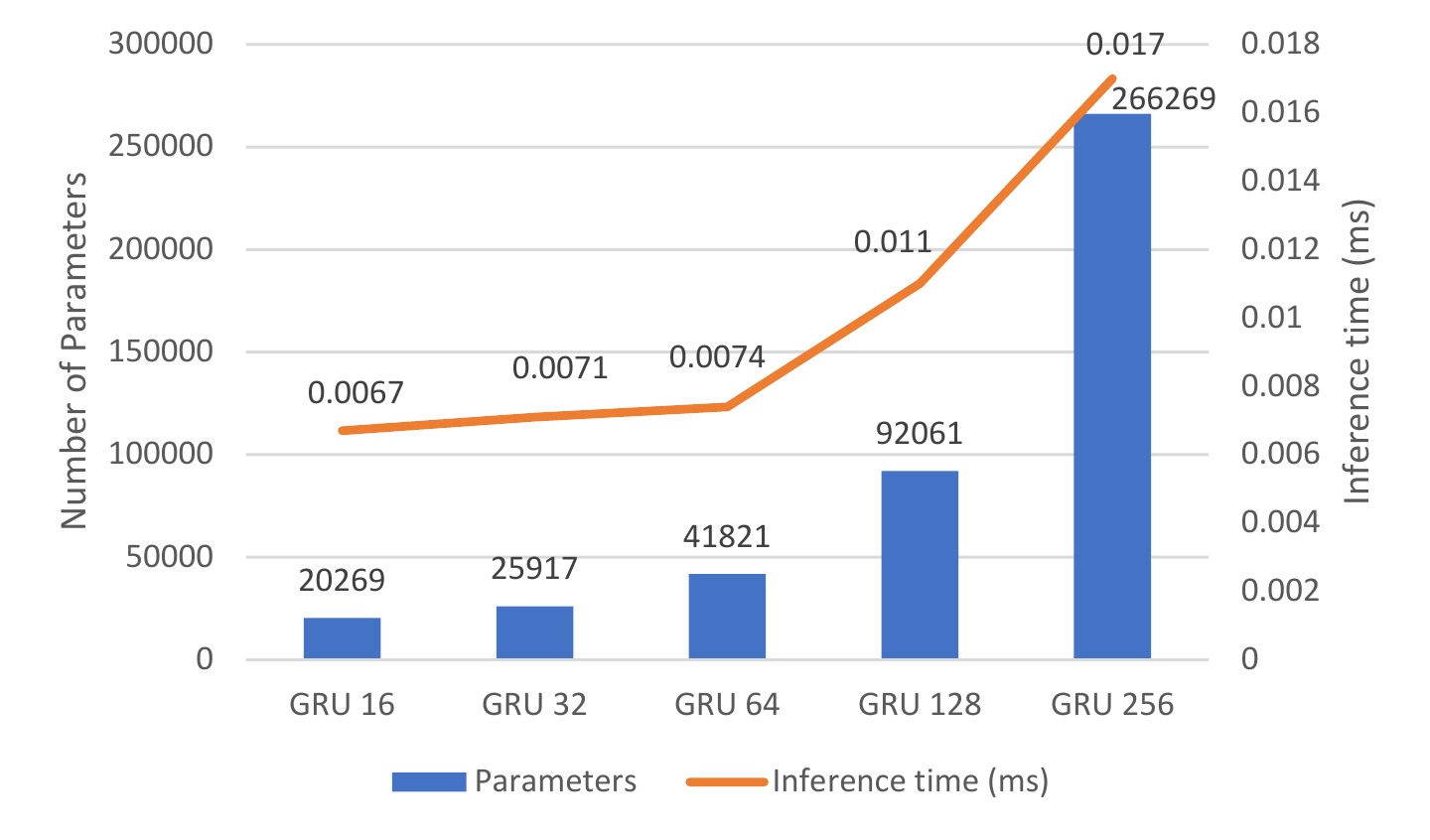}
\caption{Complexity expressed in difference of the number of hidden layers in the GRU.}
\label{Comparing GRU Hidden Layer Complexity}
\end{figure}
\begin{figure}[!t]
\centering
\includegraphics[width=3.4 in]{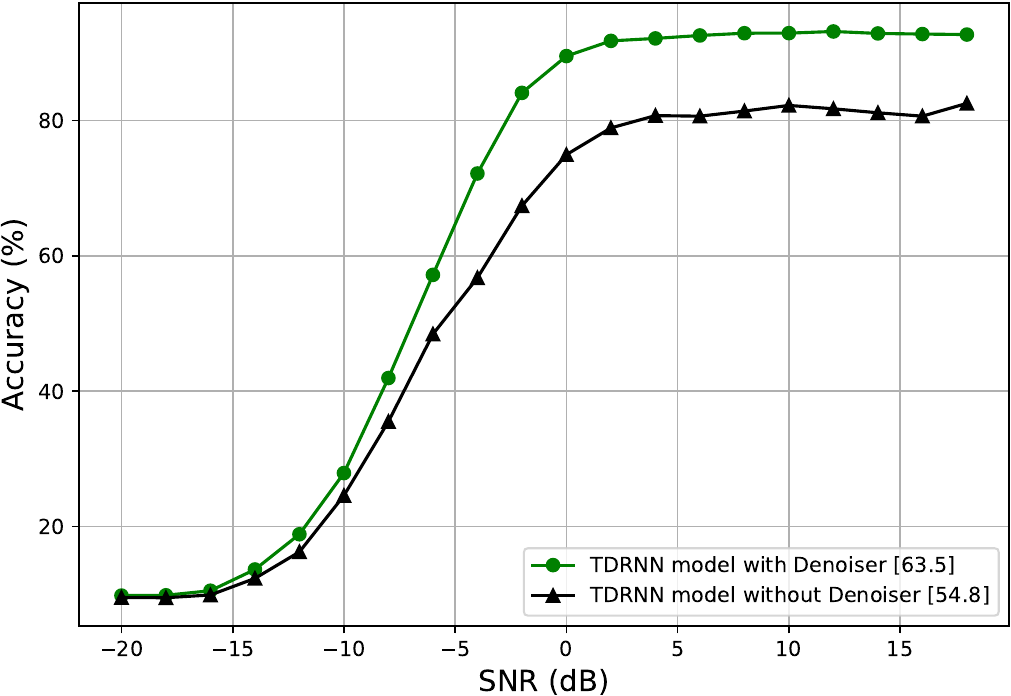}
\caption{Comparing TDRNN model with and without Threshold Denoiser.}
\label{Comparing TDRNN model with and without Threshold Denoiser}
\end{figure}

The first experiment examines the effectiveness of the hidden layers in the GRU within the complete TDRNN framework. To assess the impact of the number of GRU hidden layers, we analyze the accuracy and complexity of the TDRNN with a varying number of hidden GRU layers, specifically 16, 32, 64, 128, and 256 hidden layers. In Fig. \ref{Classification Performance of different GRU}, we present the classification accuracy versus SNR for different numbers of GRUs. The results demonstrate that as the number of hidden layers increases, there is a general improvement in accuracy allowing the network to extract more relevant information. Notably, there is an improvement of approximately 3.5\% when transitioning from GRU models with 32 hidden layers to those with 64 hidden layers, with marginal improvements achieved with fewer layers (from 16 to 32 hidden layers).

Regarding complexity, Fig. \ref{Comparing GRU Hidden Layer Complexity} presents a comparison of the TDRNN complexity expressed in terms of the number of parameters and the inference time per a single modulation example, for various numbers of hidden layers in the GRU. Overall, there is a progressive increase in complexity as the number of hidden layers rises. Examining Fig. \ref{Comparing GRU Hidden Layer Complexity}, it becomes evident that GRU models with 16, 32, and 64 hidden layers exhibit a relatively small number of parameters, ranging from 20,000 to 42,000. However, GRU models with 128 and 256 hidden layers have 92,061 and 266,269 parameters, respectively, which are approximately two and six times higher than those of the 64 hidden layer model.

In terms of inference time per a single modulation example, the fastest models are the GRU models with 16 and 32 hidden layers. The former can finish classification in 0.0067 ms and the latter at 0.0071 ms. Close behind is the GRU model with 64 hidden layers, with an inference time of 0.0074 ms. Comparatively, the slower models include the GRU models with 128 and 256 hidden layers, with inference times of 0.011 ms and 0.017 ms, respectively. These inference times are approximately 1.5 and 2.3 times lower than those of the GRU models with 64 hidden layers.

Based on this analysis, we see that selecting the GRU model with 64 hidden layers offers an excellent balance between high accuracy and computational complexity. It exhibits significant improvements in accuracy, when compared to models with fewer layers, while it maintains a relatively small number of parameters and inference time per a single modulation example.

In the second experiment, we analyzed the impact of the Threshold Denoiser on the performance of the TDRNN model. We compared two variations: one incorporating the Threshold Denoiser and one without it. The results, presented in Fig.  \ref{Comparing TDRNN model with and without Threshold Denoiser} clearly demonstrate that the TDRNN model with the Threshold Denoiser consistently outperforms the model without it across all SNR conditions. Notably, the denoiser-enhanced TDRNN model achieves a peak accuracy of 93\%, which is approximately 11\% higher than the accuracy of the model without the denoiser.

Furthermore, the Threshold Denoiser significantly improves the model’s average accuracy, increasing it by 15.8\%, from 54.8\% to 63.5\%. This substantial enhancement highlights the effectiveness of the Threshold Denoiser in mitigating noise and improving classification performance under varying SNR conditions. These findings emphasize the critical role of adaptive noise suppression in enhancing the robustness and accuracy of the TDRNN model.

For the next experiment, we examine the impact of the number of GRU layers on the overall performance of the model. Table 3 provides a comprehensive comparison of the TDRNN model performance with varying numbers of GRU layers, focusing on parameters, inference time per a single modulation example, and average accuracy. As the number of GRU layers increases, there is a corresponding increase in the number of parameters, increasing the model complexity. At the same time, there is a marginal increase in inference time per a single modulation example as more layers are added. However, it is noteworthy that the average accuracy remains consistently around 63.3\% across different layer configurations, suggesting that the influence of GRU layers on accuracy is limited. Consequently, given the trade-off between model complexity and computational efficiency, it may be sufficient to choose a model with a single GRU layer. This decision is supported by the fact that additional layers lead to a higher number of parameters and a slightly longer inference time without yielding significant improvements in average accuracy. Therefore, opting for a single GRU layer strikes a balance between accuracy and computational efficiency.

\begin{table*}[]
\centering
\caption{MODEL COMPARISON ON TWO DATASETS (A: RML2016.10A, B: RML2018.01A)}
\setlength{\tabcolsep}{12pt}
\renewcommand{\arraystretch}{1.5}
\label{compare model}
%\resizebox{1.7\columnwidth}{!}{%
\begin{tabular}{|c|c|c|c|c|l}
\cline{1-5}
Model                        & Datasets & Parameters      & Inference time(ms) & Average accuracy(\%) &  \\ \cline{1-5}
\multirow{2}{*}{MCLDNN}      & A        & 406 199         & 0.16               & 60.49                &  \\ \cline{2-5}
                             & B        & 407 876         & 0.142              & \textbf{61.68}       &  \\ \cline{1-5}
\multirow{2}{*}{PET-CGDNN}   & A        & 71 871          & 0.063              & 59.65                &  \\ \cline{2-5}
                             & B        & 75 340          & 0.083              & 60.83                &  \\ \cline{1-5}
\multirow{2}{*}{MCNet}       & A        & 90 763          & 0.021              & 58.1                 &  \\ \cline{2-5}
                             & B        & 126 616         & \textbf{0.051}     & 57.13                &  \\ \cline{1-5}
\multirow{2}{*}{SCNN}        & A        & 104 395         & 0.009              & 47.82                &  \\ \cline{2-5}
                             & B        & 2 110 040       & 0.053              & 15.19                &  \\ \cline{1-5}
\multirow{2}{*}{TDRNN(ours)} & A        & \textbf{41 821} & \textbf{0.007}     & \textbf{63.5}        &  \\ \cline{2-5}
                             & B        & \textbf{43 642} & 0.061              & 60.58                &  \\ \cline{1-5}
\end{tabular}%
%}
\end{table*}

\subsection{Comparative Evaluation}
Next, we conduct a comprehensive comparison between the TDRNN model and other recent AMC models, namely MCLDNN \cite{rw5}, PET-CGDNN \cite{rw13}, MCNet \cite{b17}, and SCNN \cite{rw9}. To ensure a fair evaluation, all systems are trained and evaluated using the same dataset and under the same experimental conditions. 
 
\begin{figure}[!t]
\centering
\includegraphics[width=3.4in]{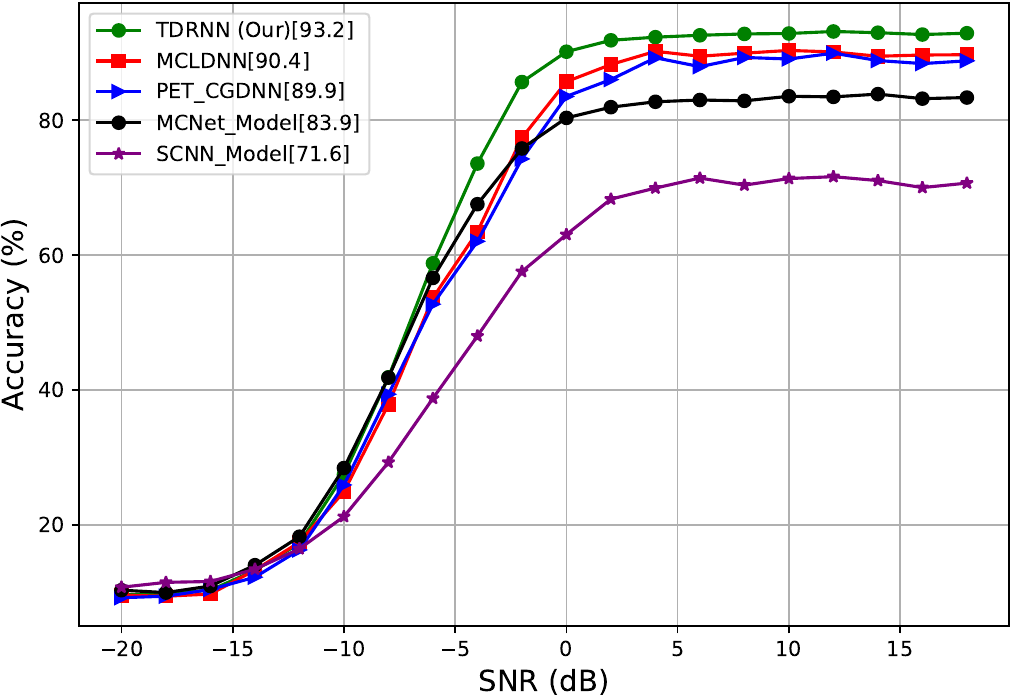}
\caption{Classification accuracy with different SNR in five different models on RadioML 2016.10A dataset.}
\label{fig7}
\end{figure}
\begin{figure}[!t]
\centering
\includegraphics[width=3.4 in]{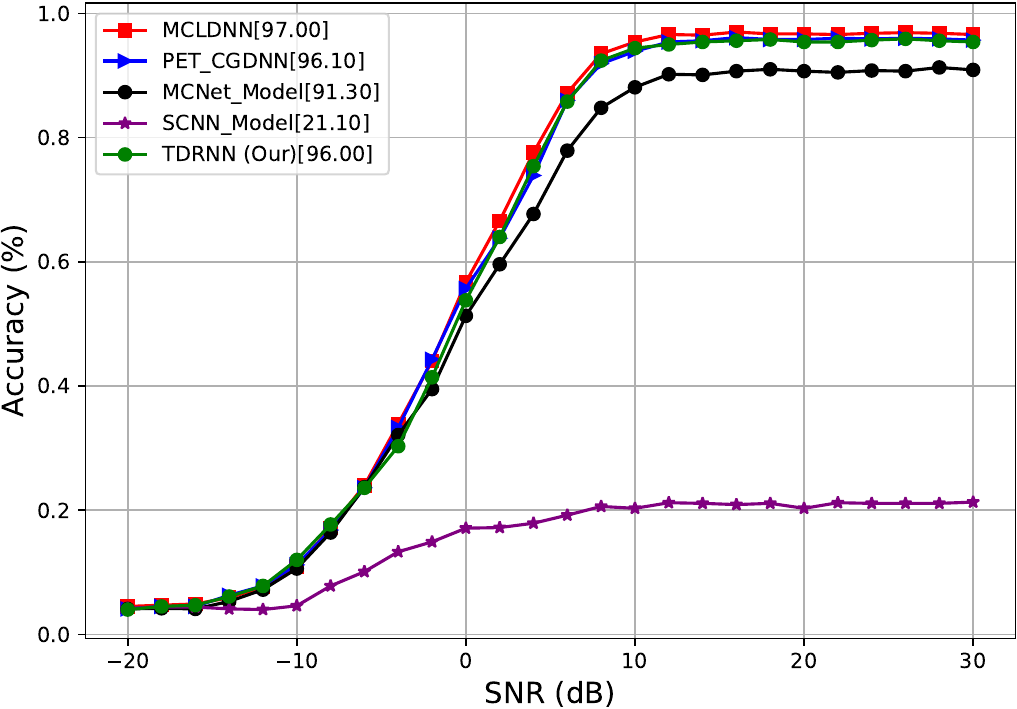}
\caption{Classification accuracy with different SNR in five different models on the RadioML 2018.01A dataset.}
\label{fig8}
\end{figure}
 First, we compare the accuracy of all five models on the RadioML 2016.10A dataset and then on the RadioML 2018.01A dataset, with the results shown in Fig. \ref{fig7} and Fig. \ref{fig8}, respectively. The results on the RadioML 2016.10A dataset demonstrate clearly that the TDRNN consistently outperforms the other models, achieving the highest accuracy values. From -8dB SNR to the highest SNR values, the TDRNN achieves a maximum accuracy of 93.2\%, followed by MCLDNN with 90.4\%, PET-CGDNN with 89.9\%, MCNet with 83.9\%, and SCNN with 71.6\%. Regarding the RadioML 2018.01A dataset, MCLDNN achieves the highest accuracy of 97\%. The PET-CGDNN and TDRNN closely follow with an almost similar highest accuracy of 96\%. On the other hand, the SCNN performs substantially poorer with a maximum accuracy of only 21\%.  Overall, for this dataset, we notice that the three schemes TDRNN, PET-CGDNN, and MCLDNN achieve almost indistinguishable accuracy. 

The highest accuracy is not the only metric we are interested in since it is not always the only metric in applications that require real-time signal classification. Consequently,  we also consider the average accuracy, the number of parameters of the model, and inference time. Table 4 provides a comprehensive comparison of these metrics for the five models. Regarding the RadioML 2016.10A dataset, it is evident from this table that the TDRNN model achieves the highest average accuracy of 63.6\%, which is approximately 3-4\% higher than MCLDNN and PET-CGDNN, and around 6\% and 16\% higher than MCNet and SCNN, respectively. These results highlight the superiority of the TDRNN model in terms of accuracy compared to the other state-of-the-art AMC models. For the RadioML 2018.01A dataset, the TDRNN model achieves a competitive accuracy of 60.58\%, trailing PET-CGDNN and MCLDNN by only 0.5-1\%.  

In terms of computational complexity, our model has the fewest parameters, approximately by a factor of two to three times as low as MCNet, PET-CGDNN, and SCNN, and approximately 10 times lower than that of MCLDNN. This is of critical importance for modern AI/DNN-enabled systems where the reduction of the number of model parameters is itself a central goal \cite{ps11, ps12}. It is worth noting that the parameters of MCLDNN, PET-CGDNN, and TDRNN exhibit slight variations when evaluated on the RadioML 2018.01A dataset, each within a margin of 5\%. However, the parameters of SCNN experience a significant increase, approximately 20-fold when testing on the the RadioML 2018.01A dataset. 

One of the most important metrics in our comparison is the inference time per a single modulation example. Looking at the results for the RadioML 2016.10A dataset, our model is the fastest, with only 0.007 ms required to produce a result, followed by the SCNN model with an inference time of 0.009 ms. The inference time per a single modulation example for TDRNN is three times lower than that of MCNet, nine times lower than PET-CGNN, and over 23 times lower than that of MLCDNN. When testing the RadioML 2018.01A dataset, our model is the second fastest with an inference time of 0.061 ms, trailing MCnet by only 0.01 ms (that achieves lower accuracy).

\subsection{Discussion}
When considering the requirements of 6G networks, both the SCNN and TDRNN models were able to achieve a low inference time, specifically 0.009 ms and 0.007 ms for the RadioML 2016.10A dataset, and 0.053 ms and 0.061 ms for RadioML 2018.01A. While both models demonstrated favorable results in terms of inference time, TDRNN achieved a significantly higher level of accuracy and worked well on both datasets. Additionally, TDRNN achieved its results with notably fewer parameters in comparison to SCNN and all the other models. As a result, TDRNN not only satisfies the expected stringent requirements of 6G networks but also enhances other aspects of system performance when compared to existing models like the complexity. Based on evaluations of RadioML 2016.10A, our model has surpassed five others in terms of accuracy, inference time, and number of parameters. Next we investigated the RadioML 2018.01A dataset, which comprises a wide range of modulations and a very large number of training examples. Traditionally, a large and complex deep learning model is used to classify this dataset, such as MCLDNN. However, our model was able to achieve a comparable level of accuracy but with a lower parameter count and a shorter inference time. This shows that our lightweight model is an efficient and accurate option for classifying both datasets. 

The presented thorough evaluation suggests that TDRNN is a step towards a more comprehensive approach for building signal classification models by considering not only the highest/average accuracy, but also the model size and inference time. Emerging architectures that will be designed in the future will need to consider all these aspects especially when targeting AMC which is an application typically deployed in real-time systems.

% In the final experiment we assess the accuracy of our proposed AMC models across various modulation types. Fig. 8 presents the confusion matrix for an SNR of 18 dB, revealing that the TDRNN model achieves high accuracy in classifying most modulation types, with the exception of WBFM and AM-DSB. TDRNN demonstrates precise predictions for 10 out of 11 modulation types, with accuracy values above 94\%, even achieving 100\% accuracy for AM-DSB, CPFSK, GFSK, and 4-PAM modulations. However, TDRNN encounters difficulty distinguishing WBFM from AM-DSB. This confusion arises due to the presence of a silent period with only carrier tones in both WBFM and AM-DSB, resulting in significant information loss within the limited observation window of the signal in the dataset.
% Furthermore, WBFM and AM-DSB have high 
% similarity in their time domain representation, as can be seen from the time domain waveforms illustrated in Fig. 9.
% This similarity further adds to the challenge faced by the AMC models in accurately distinguishing between these two modulations. The shared characteristics in their time domain waveforms make it difficult for the models to discern the subtle differences between WBFM and AM-DSB.
\section{CONCLUSION}
In this study, we introduced TDRNN which is a new approach for AMC that is aligned with the expected KPIs of 6G systems in terms of latency. TDRNN combines denoising techniques and RNN-based classification for enhanced modulation classification on three metrics. By using a Threshold Denoiser (TD) to eliminate noise and irrelevant features, we achieve almost a 16\% accuracy improvement when compared to models without TD. Our proposed TDRNN system demonstrates significant performance improvements over existing approaches: compared to MCLDNN, it achieves a 4.98\% improvement in accuracy while exhibiting 23 times faster inference times and requires ten times fewer parameters on the RadioML 2016.10A dataset. 
For the more comprehensive RadioML 2018.01A dataset TDRNN offers by far the smallest model size, with similar inference time and average accuracy with other models. Our proposed system demonstrates promising applicability in 6G networks, where ultra-low latency and high throughput are imperative. This is demonstrated by its remarkable inference times of 0.007 ms and 0.061 ms across both datasets.

\bibliographystyle{IEEEtran}
\bibliography{Bib1}

\begin{IEEEbiography}[{\includegraphics[width=1in,height=1.25in,clip,keepaspectratio]{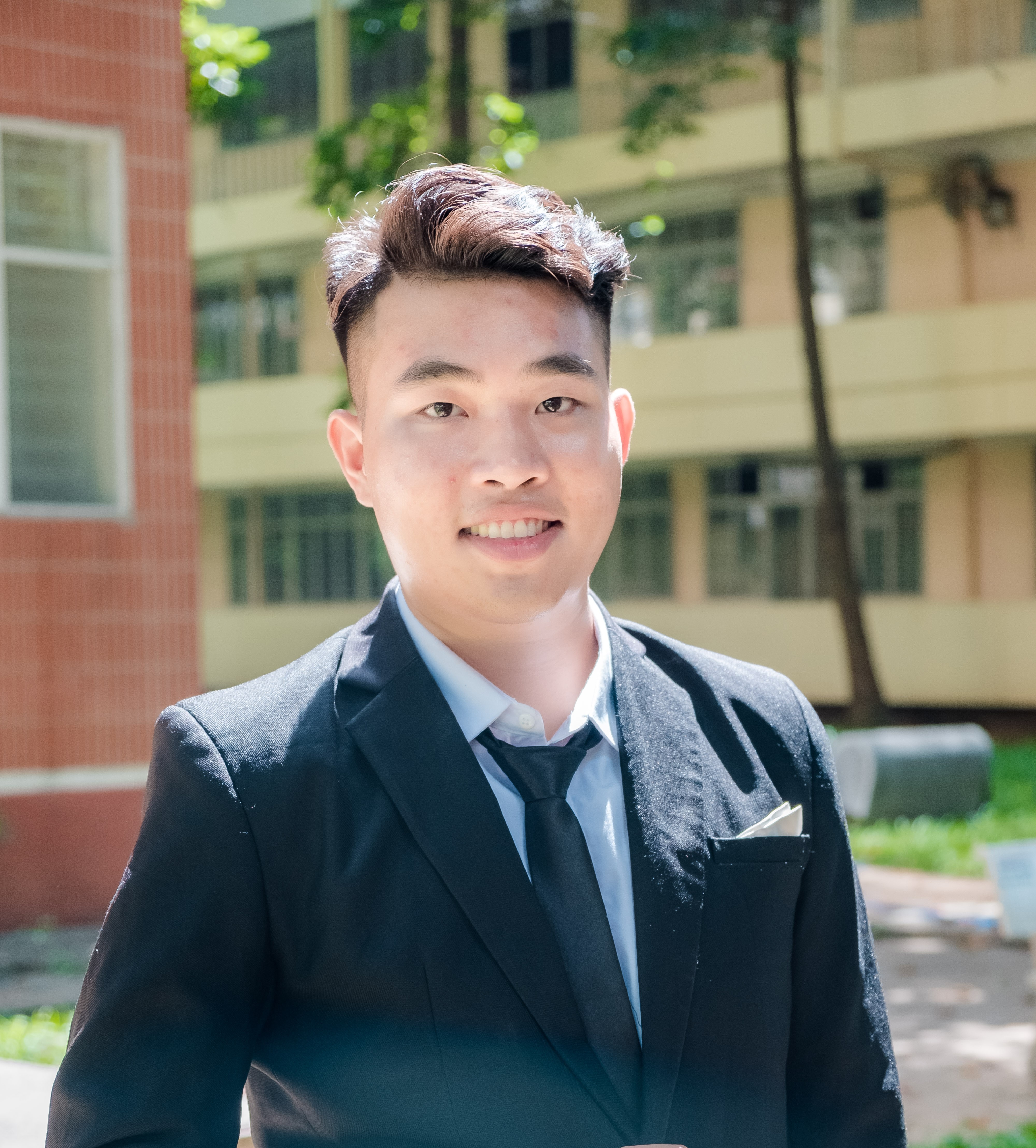}}]{To Truong An }(Student Member, IEEE) received a B.S. degree in mechanical engineering from the Ho Chi Minh City University of Technology and Education, Vietnam, in 2016. He then obtained a Master’s degrees in Intelligent Mechatronics Engineering and Convergence Engineering for Intelligent Drone from Sejong University, Seoul, South Korea. Currently, he is pursuing a Ph.D. in the School of Electronics, Electrical Engineering and Computer Science at Queen’s University Belfast, UK. He had a two-year industry background as a Research Engineer at Samsung Electronics, Ho Chi Minh, Vietnam, in 2020 and 2021. His research interests include wireless communication, sixth-generation, network security, physical layer security, automatic modulation classification, machine learning.
\end{IEEEbiography}
\vspace{-2\baselineskip}
\begin{IEEEbiography}[{\includegraphics[width=1in,height=1.25in,clip,keepaspectratio]{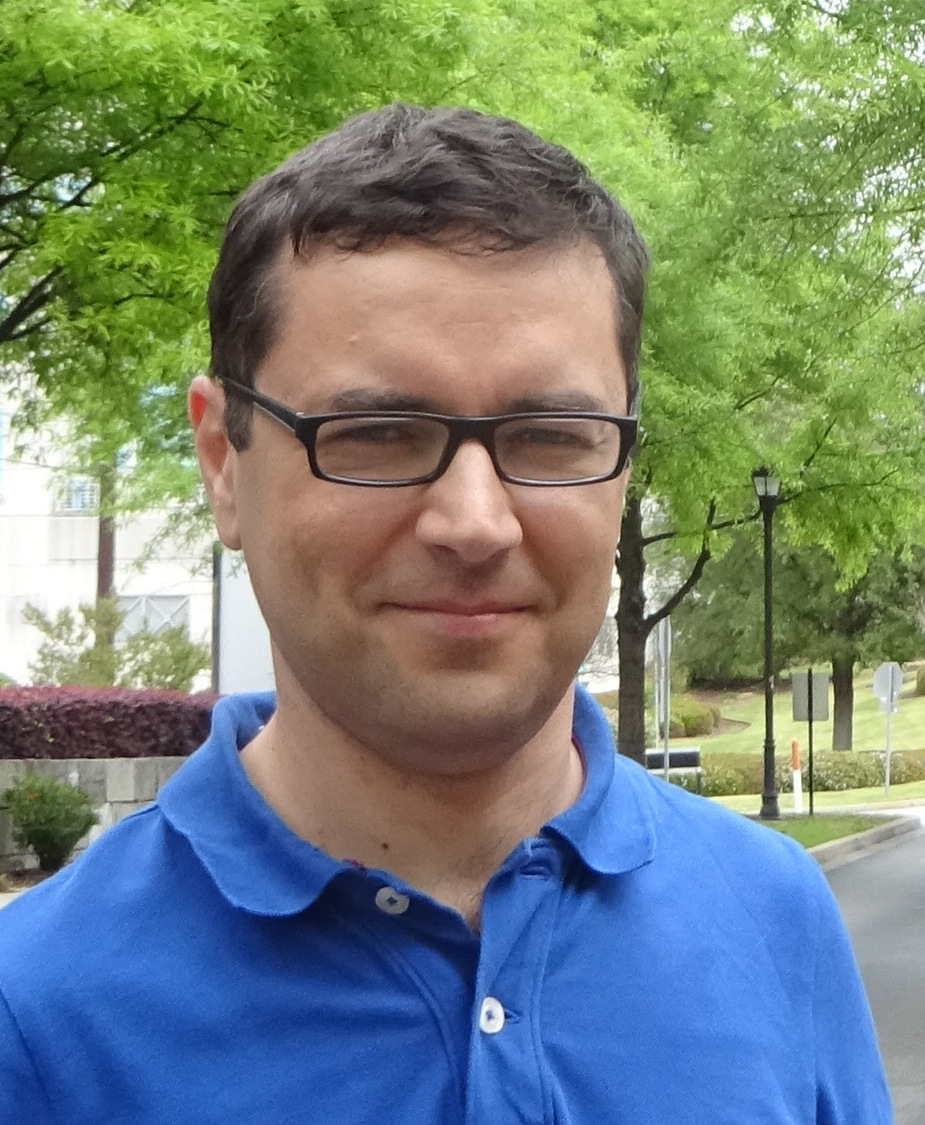}}]{Antonios Argyriou } received the Diploma in electrical and computer engineering from Democritus University of Thrace, Greece, and the M.S. and Ph.D. degrees in electrical and computer engineering as a Fulbright scholar from the Georgia Institute of Technology, Atlanta, USA. Currently, he is a Professor at the department of electrical and computer engineering, University of Thessaly, Greece. In the past he has been an Assistant Professor at the University of South Carolina, a Senior Research Scientist at Philips Research, The Netherlands, and a Senior Engineer at Soft.Networks, Atlanta. Dr. Argyriou has served as guest editor for the IEEE Transactions on Multimedia Special Issue on Quality-Driven Cross-Layer Design, and he was also a lead guest editor for the Journal of Communications, Special Issue on Network Coding and Applications. Dr. Argyriou serves in the TPC of several international conferences and workshops. His research expertise is in the areas of wireless communications, RADAR systems, machine learning, and statistical signal processing theory and applications.
\end{IEEEbiography}
\vspace{-2\baselineskip}
\begin{IEEEbiography}[{\includegraphics[width=1in,height=1.3in,clip,keepaspectratio]{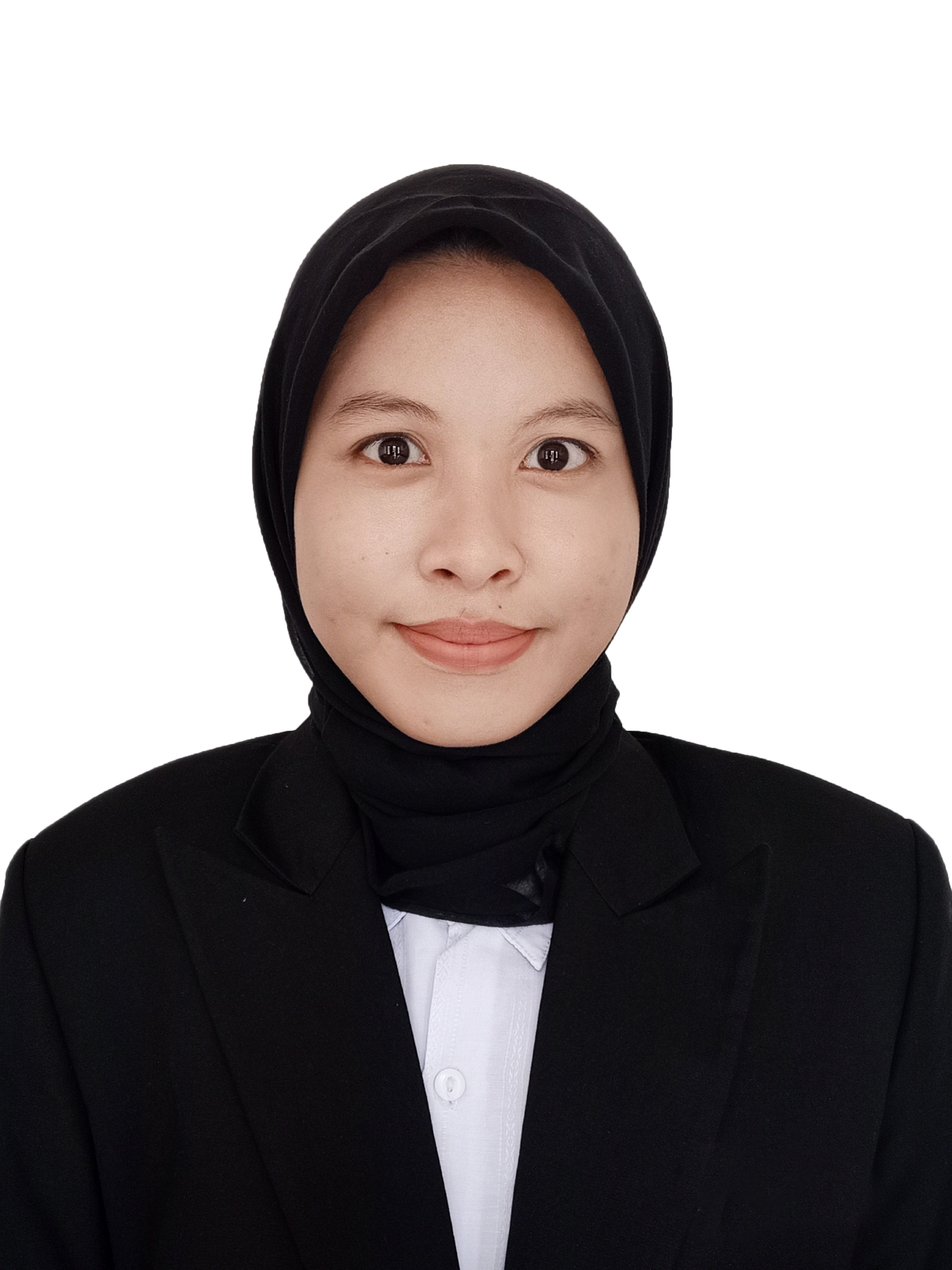}}]{Annisa Anggun Puspitasari }
received her B.E. degree in telecommunication engineering from the Electronics Engineering Polytechnic Institute of Surabaya, Indonesia, in 2021. She later obtained Master’s degrees in Intelligent Mechatronics Engineering and Convergence Engineering for Intelligent Drone from Sejong University, Seoul, South Korea. Currently, she is pursuing a Ph.D. in the AI Robotics department, Sejong University, Seoul, South Korea. 

From Oct 2021 to June 2022, she was an adjunct researcher at Okayama University, Japan. Her research interests include wireless communication systems, signal processing, machine learning, and quantum-based machine learning applications.
\end{IEEEbiography}
\vspace{-2\baselineskip}

\begin{IEEEbiography}[{\includegraphics[width=1.5in,height=1.3in,clip,keepaspectratio]{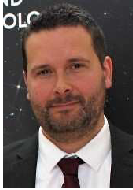}}]{Simon L. Cotton }(Fellow, IEEE)
received the B.Eng. degree in electronics and software from Ulster University, Ulster, U.K., in 2004, and the Ph.D. degree in electrical and electronic engineering from Queen’s University Belfast, Belfast, U.K., in 2007. He was a Research Fellow (2007–2011), a Senior Research Fellow (2011–2012), a Lecturer and an Assistant Professor (2012–2015), and a Reader and an Associate Professor (2015–2019) with Queen’s University Belfast, where he is currently a Full Professor and the Director of the Centre for Wireless Innovation (CWI). He has authored or coauthored over 150 publications in major IEEE/IET journals and refereed international conferences, three book chapters, and two patents. His research interests include propagation measurements and statistical channel characterization. His other research interests include cellular device-to-device, vehicular, and body-centric communications. He received the H. A. Wheeler Prize in July 2010 from the IEEE Antennas and Propagation Society, for the best applications journal paper in the IEEE Transactions on Antennas and Propagation in 2009; and the Sir George Macfarlane Award from the U.K. Royal Academy of Engineering in July 2011, in recognition of his technical and scientific attainment, since graduating from his first degree in engineering.

\end{IEEEbiography}

\begin{IEEEbiography}[{\includegraphics[width=1in,height=1.25in,clip,keepaspectratio]{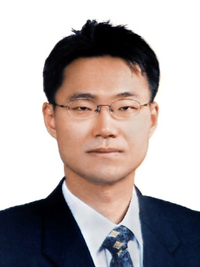}}]{Byung Moo Lee }(Senior Member, IEEE) 
received the Ph.D. degree in electrical and computer engineering from the University of California at Irvine, CA, USA, in 2006.

He had ten years of industry experience including, research positions at the Samsung Electronics Seoul R\&D Center, Samsung Advanced Institute of Technology (SAIT), and the Korea Telecom (KT) R\&D Center. He is currently a Professor at the Department of Artificial Intelligence and Robotics, Sejong University, Seoul, South Korea. During his industry experience, he participated in IEEE 802.16/11, Wi-Fi Alliance, and 3GPP LTE standardizations, and also participated at the Mobile VCE and the Green Touch Research Consortiums, where he made numerous contributions and filed a number of related patents. His research interests include wireless communications, signal processing, and machine learning applications. 

Dr. Lee was the Vice Chairperson of the Wi-Fi Alliance Display MTG, from 2015 to 2016.
\end{IEEEbiography}

\vfill

\end{document}